\documentclass[letter]{aa}
\usepackage{txfonts}
\usepackage{graphicx}
\usepackage{xcolor}

\def\H2{H$_2$}

\begin{document}

\title{Water and acetaldehyde in HH212: The first hot corino in Orion} 
\author{C. Codella \inst{1,2,3} 
 \and 
C. Ceccarelli \inst{2,3,1} \and
S. Cabrit \inst{4,2,3} \and
F. Gueth  \inst{5} \and
L. Podio \inst{1} \and
R. Bachiller \inst{6} \and
F. Fontani \inst{1} \and
A. Gusdorf \inst{4} \and
B. Lefloch \inst{2,3} \and
S. Leurini \inst{7} \and
M. Tafalla \inst{6} } 

\institute{
INAF, Osservatorio Astrofisico di Arcetri, Largo E. Fermi 5,
50125 Firenze, Italy
\and
Univ. Grenoble Alpes, Institut de
Plan\'etologie et d'Astrophysique de Grenoble (IPAG), 38401 Grenoble, France
\and
CNRS, Institut de
Plan\'etologie et d'Astrophysique de Grenoble (IPAG), 38401 Grenoble, France
\and
LERMA, UMR 8112 du CNRS, Observatoire de Paris, \'Ecole  Normale Sup\'erieure, 61 Av. de l'Observatoire, 75014
Paris, France
\and
IRAM, 300 rue de la Piscine, 38406 Saint Martin d'H\`eres, France
\and
IGN, Observatorio Astron\'omico Nacional, Alfonso XII 3, 28014, Madrid, Spain
\and
Max-Planck-Institut f\"ur Radioastronomie, Auf dem H\"ugel 69, 53121 Bonn, Germany
}

\offprints{C. Codella, \email{codella@arcetri.astro.it}}
\date{Received date; accepted date}

\authorrunning{Codella et al.}
\titlerunning{Water and acetaldehyde around HH212}

\abstract
{}
{Using the unprecedented combination of 
high resolution and sensitivity offered by ALMA, 
we aim to investigate whether and how hot corinos, circumstellar disks, 
and ejected gas are related in young solar-mass protostars.} 
{We observed CH$_3$CHO and deuterated water (HDO) high-excitation ($E_{\rm u}$ up to 335 K) 
lines towards the Sun-like protostar HH212--MM1.} 
{For the first time, we have obtained images  
of CH$_3$CHO and HDO emission in 
the inner $\simeq$ 100 AU of HH212.
The multifrequency line analysis allows us to contrain 
the density ($\geq$ 10$^{7}$ cm$^{-3}$),  
temperature ($\simeq$ 100 K), and CH$_3$CHO abundance ($\simeq$ 0.2--2 $\times$ 10$^{-9}$) 
of the emitting region.
The HDO profile is asymmetric at low velocities ($\leq$ 2 km s$^{-1}$ from $V_{\rm sys}$). 
If the HDO line is optically thick, this points to an extremely small ($\sim$ 20--40 AU)
and dense ($\ge$ 10$^{9}$ cm$^{-3}$) emitting region.}
{We report the first detection of a hot corino
in Orion. The HDO asymmetric profile indicates a contribution of outflowing gas from the 
compact central region, possibly associated with a dense disk wind.}


\maketitle

\section{Introduction: The HH212 laboratory}

The birth of a Sun-like star is a complex game played by several
participants 
whose respective roles are not yet entirely clear.  On the one
hand, the star-to-be accretes matter from a collapsing envelope. 
It is commonly believed that the
gravitational energy released in the process 
heats up the material surrounding the protostar, 
creating warm regions ($\sim 100$ K)  enriched by complex
organic molecules (COMs) called hot corinos (e.g. Ceccarelli et al. 2007).
On the other hand, 
the presence of angular momentum and magnetic fields 
leads to two consequences: 
(i) the formation of circumstellar
disks, also called protoplanetary disks, and (ii) 
substantial episodes of matter ejection (e.g. Frank et al. 2014,
and references therein). 

Despite the progress achieved 
in the last decade, to our knowledge
only three hot corinos have been  imaged so far
in COMs on $\leq$ 100 AU scale (IRAS16293-2422, NGC1333-IRAS2A, and IRAS4A;
e.g. 
J\o{}rgensen et al. 2012; Maury et al. 2014; Taquet et al. 2015,
and references therein). Hence, several 
questions about these three components (hot corino, circumstellar disk,
and ejected material) are still unanswered.  
What is the origin and composition of the hot corinos: are they thermally or
shocked heated regions? What is the origin of the ejections:
are they due to disk or stellar winds? How are these three phenomena linked? 
In addition to the physical and evolutionary connection, the three
phenomena have one thing in common:  they all need to be studied using 
mm observations at high spatial resolution (on scales 
$\leq$ 100 AU). 

HH212 is located in Orion (at 450 pc) and is one of the best laboratories in which to study the 
(Class 0) protostellar stage. The HH212--MM1 protostar is hidden in a rotating and infalling envelope
(Wiseman et al. 2001; Lee et al. 2014) and drives a spectacular jet and outflow. 
It has 
been extensively studied 
with the SMA and IRAM PdBI on scales 
down to $\simeq$ 0$\farcs$3--0$\farcs$4 (Lee et al. 2006, 2007, 2008; Codella et al. 2007; Cabrit et al. 2007, 2012),
showing a microjet with inner peaks at $\pm$1--2$\arcsec$ = 450--900 AU of the protostar.
Further observations performed with ALMA Early Science in Band 7  
in HCO$^+$, C$^{17}$O, and SO indicate a small-scale velocity gradient 
along the equatorial plane consistent with a rotating disk of $\simeq 0\farcs2$ = 90 AU
around a $\simeq 0.3\pm0.1 M_{\rm \odot}$ source (Lee et al. 2014; Codella et al. 2014; Podio et al. 2015).
The HH212 region is thus, so far, the only protostar 
where both a bright 
bipolar jet and a compact rotating disk have been
revealed. Given its association with all the ingredients of the
Sun-like star formation recipe, HH212 stands out as being the perfect target 
for investigating the link between hot corino, disk, and outflow in the protostellar stage. 

In this letter, we 
further exploit the ALMA dataset of Codella et al. (2014) 
to probe the inner
region of HH212. Specifically, we 
report the detection of twelve high-lying ($E_{\rm u} \geq 150$ K) lines from
acetaldehyde (CH$_3$CHO) and one ($E_{\rm u}$ = 335 K) line from
deuterated water (HDO). They reveal the presence of a hot corino previously
unknown in this source and, possibly, probe the base of 
a slow disk wind.

\section{Observations} 

HH212 was observed 
at 850 $\mu$m with ALMA using 24 12 m antennas  
on 2012 December 1 (Early Science Cycle 0 phase; Codella et al. 2014). 
The baselines were between 20 m and 360 m with 
a maximum unfiltered scale of 3$\arcsec$.
The spectral windows 333.7--337.4 GHz and 345.6--349.3 GHz 
were observed using spectral units of 488 kHz (0.42--0.44 km s$^{-1}$).
Calibration was carried out following standard procedures, 
using quasars J0538--440, J0607--085, Callisto, and Ganymede.
Spectral line imaging was achieved with the CASA package, while 
data analysis was performed using the 
GILDAS\footnote{http://www.iram.fr/IRAMFR/GILDAS} package.
The continuum-subtracted images have a typical 
clean-beam FWHM of $0\farcs65\times0\farcs47$ 
(PA = 35$\degr$), and an rms noise level of
3--4 mJy beam$^{-1}$ in 0.44 km s$^{-1}$ channels.  
Positions are given with respect to the MM1 protostar continuum peak
located at $\alpha({\rm J2000})$ = 05$^h$ 43$^m$ 51$\fs$41, 
$\delta({\rm J2000})$ = --01$\degr$ 02$\arcmin$ 53$\farcs$17
(Lee et al. 2014).

\section{Results and discussion}

The ALMA 8 GHz bandwidth presents  
a very rich spectrum with many lines in emission 
towards the MM1 protostar (Fig. 1) that have revealed a number of  
high-excitation molecular lines (see also Table 1), among which
(i) twelve lines of acetaldehyde (CH$_3$CHO) with  
$E_{\rm u}$ between 150 K and 200 K 
and (ii) the 3$_{3,1}$--4$_{2,2}$ line 
of deuterated water (HDO) from $E_{\rm u}$ = 335 K (see Figs. 2 and 3).  
The lines were identified using the Jet 
Propulsion Laboratory (JPL) molecular database (Pickett et al. 1998).
Figure 1  
shows that CH$_3$CHO and HDO are only observed  towards the
protostellar position and are spatially unresolved.
For the first time these high-excitation molecular lines have been detected
towards the HH212 inner region revealing the existence of hot gas
around the MM1 protostar driving the bipolar SiO jet.
The line profiles peak
in the +1.2,+2.0 km s$^{-1}$ range, i.e. close to
the systemic velocity\footnote{In the literature,
values between +1.3 and +1.8 km s$^{-1}$ are reported for the $V_{\rm sys}$ of HH212.
The value of +1.3 km s$^{-1}$ adopted in Codella et al. (2014) was
affected by a shift in the frequency scale; 
we adopt here +1.7 km s$^{-1}$ (Lee et al. 2014). 
However,  the results  
are not all dependent on this choice.} $V_{\rm sys}$,
which is  also quite broad (FWHM $\simeq$ 5--6 km s$^{-1}$).
We now examine the constraints brought by CH$_3$CHO and HDO 
on the nature and physical/chemical conditions of the emission region(s).

\subsection{CH$_3$CHO emission}

The CH$_3$CHO emission is symmetric around the systemic velocity and
can be explained by assuming that it originates in
the hot corino, namely the region where the dust
temperature is high enough ($\geq$ 100 K) to sublimate the frozen ice
mantles. 
Given that the CH$_3$CHO emission is spatially unresolved,
we assume a source size equal to
half a beam, namely 0$\farcs$3.
An LTE (Local Thermodynamic Equilibrium) optically thin analysis of
CH$_3$CHO (using only lines with
less than 30\% contamination from blended lines, in terms
of integrated emission, see Table 1) indicates 
temperature of 87$\pm$47 K and a column density of 2$\pm$1 $\times$ 10$^{15}$
cm$^{-2}$ (see the the rotational diagram in Fig. A.1 and the synthetic spectra in Fig. 2).
With these values the opacity is less than 0.4. 
In the case of a smaller emitting size, e.g. 0$\farcs$06 (as assumed for HDO in
Sect. 3), $N_{\rm CH_3CHO}$ increases by a factor of $\sim$ 20. 

Unfortunately, we cannot use the continuum emission to estimate the
H$_2$ column density (and hence the CH$_3$CHO abundance)
because the submillimetre continuum peak seen in
interferometric maps is optically thick
(Codella et al. 2007; Lee et al. 2008, 2014).
Taquet et al. (2015) measured N(H$_2$) $\simeq$ 10$^{24}$--10$^{25}$ cm$^{-2}$
towards the NGC1333-IRAS2A and IRAS4A hot corinos on a spatial scale of $\sim$ 470
AU, i.e. a factor $\sim$ 3 higher than is sampled here. 
If we take these values and 0$\farcs$3 as emitting size, then 
$X_{\rm CH_3CHO}$ $\sim$ 0.2--2 $\times$ 10$^{-9}$, in agreement with 
the value recently derived in the IRAS16293-2422 hot corino (using single-dish data; 3 $\times$
10$^{-9}$: Jaber et al. 2014).

\begin{table}
  \caption{List of unblended transitions
detected towards HH212--MM1 and used
for the CH$_3$CHO (LTE) and HDO (LVG) analysis.}
  \begin{tabular}{lccccc}
  \hline
\multicolumn{1}{c}{Transition} &
\multicolumn{1}{c}{$\nu$$^{\rm a}$} &
\multicolumn{1}{c}{$E_{\rm u}$$^a$} &
\multicolumn{1}{c}{$S\mu^2$$^a$} &
\multicolumn{1}{c}{rms$^b$} &
\multicolumn{1}{c}{$F_{\rm int}$$^b$} \\
\multicolumn{1}{c}{($J_{\rm Ka,Kc}$)} &
\multicolumn{1}{c}{(MHz)} &
\multicolumn{1}{c}{(K)} &
\multicolumn{1}{c}{(D$^2$)} &
\multicolumn{1}{c}{(mK)} &
\multicolumn{1}{c}{(K km s$^{-1}$)} \\
\hline
\multicolumn{6}{c}{CH$_3$CHO} \\
\hline
18$_{\rm 1,18}$--17$_{\rm 1,17}$ E  & 333941.4 & 155 & 226.8 & 51 & 2.8(0.5) \\
17$_{\rm 2,15}$--16$_{\rm 2,14}$ A  & 334931.4 & 153 & 212.1 & 57 & 2.6(0.1) \\
18$_{\rm 0,18}$--17$_{\rm 0,17}$ A  & 335358.7 & 154 & 226.8 & 52 & 2.3(0.5) \\
18$_{\rm 3,16}$--17$_{\rm 3,15}$ A  & 347519.2 & 179 & 221.3 & 68 & 2.4(0.2) \\
18$_{\rm 3,16}$--17$_{\rm 3,15}$ E  & 347563.3 & 179 & 221.1 & 68 & 1.7(0.3) \\
\hline
\multicolumn{6}{c}{HDO} \\
\hline
3$_{3,1}$--4$_{2,2}$ & 335395.5 & 335 & 0.4 & 57 & 4.9(0.2) \\
\hline
\end{tabular}

$^a$ From the JPL database (Pickett et al. 1998).
$^b$ In $T_{\rm B}$ scale. \\
\end{table}

\begin{figure}
\centerline{\includegraphics[angle=0,width=4.4cm]{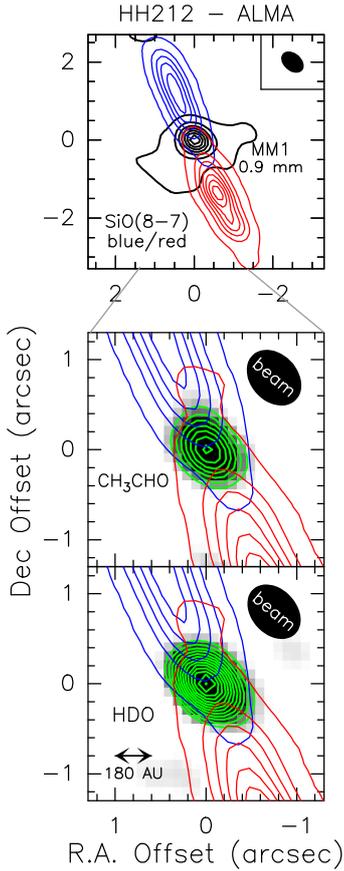}}
\caption{
{\it Upper panel:} The HH212 protostellar system
as observed by ALMA--Band 7 (Codella et al. 2014).
Blue/red contours plot the blue-/redshifted SiO(8--7) jet
at $\pm$ 8 km s$^{-1}$ from $V_{\rm sys}$,
overlaid on the 0.9 mm continuum (black contours).
Positions are with respect to the coordinates
reported in Sect. 2.
The filled ellipse shows
the synthesised beam (HPBW) for SiO:
$0\farcs63\times0\farcs46$ (PA = 49$\degr$).
{\it Middle panel:} Zoom-in of the central region:
the CH$_3$CHO(18$_{0,18}$--17$_{0,17}$)A 
emission integrated over $\pm$ 5 km s$^{-1}$ 
(green contours and grey scale) overlaid on  the SiO jet.
First contours and steps are 5$\sigma$
(50 mJy beam $^{-1}$ km s$^{-1}$) and 2$\sigma$, respectively.
The HPBW 
is $0\farcs69\times0\farcs52$ (PA = 42$\degr$).
{\it Bottom panel:}
Same as {\it middle panel} for HDO.
The HPBW 
is $0\farcs68\times0\farcs51$ (PA = 42$\degr$).}
\label{maps}
\end{figure}

\begin{figure}
\centerline{\includegraphics[angle=0,width=9cm]{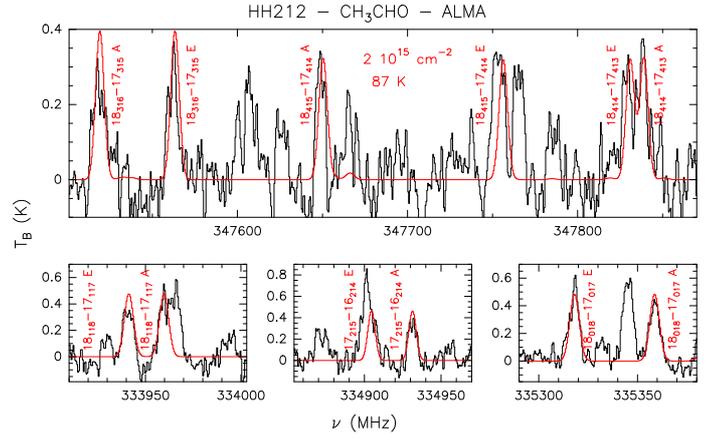}}
\caption{CH$_3$CHO emission (in $T_{\rm B}$ scale) extracted at 
the MM1 position.
The four panels show the frequency intervals where the CH$_3$CHO lines
are located (see Table 1 for those unblended with other lines).
The red line shows the synthetic spectrum which 
best reproduces the observations assuming LTE and optically thin emission 
obtained with the GILDAS--Weeds package (Maret et al. 2011) with
source size = 0$\farcs$3, $T_{\rm rot}$ = 87 K, $N_{\rm CH_3CHO}$ = 2 $\times$ 10$^{15}$ cm$^{-2}$,
FWHM linewidth = 5.0 km s$^{-1}$, and LSR velocity = +1.7 km s$^{-1}$ (see footnote 2).}
\label{ace}
\end{figure}

\subsection{HDO emission}

\begin{figure}
\centerline{\includegraphics[angle=0,width=6cm]{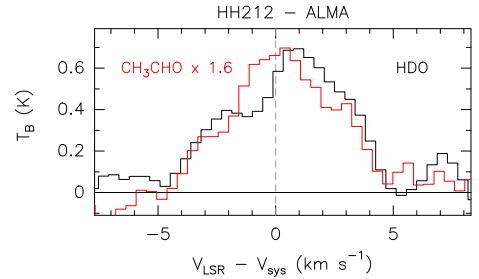}}
\caption{Comparison between HDO(3$_{3,1}$--4$_{2,2}$) 
(black line) and CH$_3$CHO(18$_{0,18}$--17$_{0,17}$)A (red, multiplied by a factor of
1.6) as observed towards HH212--MM1 (in $T_{\rm B}$, scale). The vertical dashed line 
defines the velocity with respect to $V_{\rm sys}$ = +1.7 km s$^{-1}$ (see footnote 2).}
\label{spectra}
\end{figure}

Figure 3 compares the HDO profile derived towards MM1 with that of
CH$_3$CHO(18$_{0,18}$--17$_{0,17}$)A; 
we chose this transition as representative of 
the acetaldehyde lines suffering no contamination. 
At low velocity ($\leq$ 2 km s$^{-1}$ from $V_{\rm sys}$) 
the HDO profile shows an asymmetry; the redshifted
emission is clearly brighter than the blueshifted. 
The line profile is determined by the kinematics and, in case of high optical thickness, 
by radiative effects such as self-absorption.
In order to understand the relative importance of each, 
we study two limit situations: (i) the optically thin case where the kinematics will dominate 
and (ii) the optically thick case 
where the profile will be strongly affected by radiative effects. 
To this end, we compare the observed emission with that predicted by a non-LTE LVG 
(Large Velocity Gradient) model 
using the code by Ceccarelli et al. (2003), a plane parallel 
geometry, the
collisional coefficients for the system HDO-H$_2$ computed by Faure et
al. (2012)\footnote{The collisional coefficients are extracted from
  the BASECOL database, Dubernet et al. (2013).}, and assuming a
Boltzmann distribution for the ortho-to-para H$_2$ ratio\footnote{The collisional coefficients with ortho-H2 
are a factor of 5 larger than the corresponding rate coefficients with para-H2 (Faure et al. 2012) 
at low temperatures ($\ll$ 45 K)  (Faure et al. 2012), while they are 
similar at higher temperatures.}. 
Figure 4 shows  the temperature  versus the HDO column density   
predicted to generate the observed  
signal (see Table 1). 
Given that the HDO emission is spatially unresolved, 
the upper panel shows the case of a source size equal to 
half a beam (0$\farcs$3) and  the lower panel shows a case
with a size that is lower by a  factor of five (see Sect. 3.2.2).
The plots cover densities from 10$^7$ to $10^{10}$ cm$^{-3}$.
The high-density end, $10^{10}$ cm$^{-3}$, represents LTE conditions.
The low end, 10$^7$ cm$^{-3}$, is set by the constraint that the HDO abundance
has to be smaller than 10\% of the D/H elemental abundance 
(in agreement with previous observations of HDO/H$_2$O towards hot
corinos, assuming an initial H$_2$O abundance 
of 3 $\times$ 10$^{-5}$; see Taquet et al. 2014, 2015), 
namely
$N$(HDO)/$N$(H$_2$) $\leq 3.2 \times 10^{-6}$, where $N$(H$_2$) is
consistently derived from the density and size of the source. For a
given size, this provides a limit to the permitted density-$N$(HDO)
values. In particular, for a size of 0$\farcs$3, if the density is 10$^6$
cm$^{-3}$, then $N$(HDO) should be $\leq 7 \times 10^{15}$ cm$^{-2}$. 
We find that it is not possible, then, to reproduce the HDO 
line intensity and we rule out a density of 10$^6$ cm$^{-3}$.
On the other hand, at 10$^7$
cm$^{-3}$, $N$(HDO) has to be $\leq 7 \times 10^{16}$ cm$^{-2}$,
and the temperature has to be larger than 230 K (see Fig. 4).
Armed with these predictions, we now discuss the implications
of each assumption, the optically thin and thick HDO line. 

\subsubsection{Optically thin case}

If the line is optically thin, then the observed HDO asymmetry at low velocities 
must be caused by 
intrinsic brightness asymmetries between the
red- and blueshifted components in the beam. We examine various explanations in turn. 
 
First, the HDO emission might probe the rotating disk, whose density is
expected to be larger than $\sim 10^7$ cm$^{-3}$ (Lee et al. 2014)
and whose C$^{17}$O and SO emissions also extend to $\pm$ 5 km s$^{-1}$ 
(Codella et al. 2014; Podio et al. 2015).
However, the low-velocity range of the HDO line 
being due to the protostellar disk is not
supported by the large (factor of 2) difference between (i) the  
blue and red peak brightnesses
and (ii) the 
blue and red peak velocities ($V_{\rm LSR}$--$V_{\rm sys}$ at $\leq$ +1 and $\sim$ --2 km s$^{-1}$). 
This would require a high degree of non-axisymmetry in the molecular disk emission, which has never been seen at this level
in the profiles of younger disks, not even when tidally
disturbed 
(see e.g. the RW Aur disk profile in Fig. 4 of Cabrit et al. 2006), although 
it has been revealed in some evolved {transitional} 
disks with prominent dust traps (e.g. Casassus et al. 2013).

We can also rule out an origin from the rotating infalling envelope;
infall motions of 1 km s$^{-1}$ arise from radii 0$\farcs$3 (Lee et al. 2014),
which would lead to spatially resolved peak emission, unlike the observed emission.
In addition, in this case, there is no reason for
the redshifted emission being definitely brighter than the blue.

The last and most likely possibility is that the low-velocity HDO emission 
is dominated by outflowing motions. 
The SiO(8--7) profile 
towards MM1 
derived from the same ALMA dataset (Podio et al. 2015) is definitely too broad  
(up to $\pm$ 20 km s$^{-1}$) 
for HDO to probe
the inner portion of the fast jet traced by SiO (Fig. 1). On the other hand, 
channel maps of SO 
$10_{11}-10_{10}$ ($E_{\rm u}$ = 143 K) 
have clearly revealed a low-velocity 
compact bipolar outflow (Podio et al. 2015) in the 
$\pm$ 2 km s$^{-1}$ ($\le$ 25 km s$^{-1}$ after deprojection) range.
Hence, it seems quite plausible that the low-velocity HDO  
would trace the inner region ($\leq$ 0$\farcs$3, i.e. 138 AU) of a more extended 
outflow
heated by the protostar at a
temperature high enough (at least 100 K) to release icy dust mantles 
and with a  brighter  HDO line in its red lobe. 

\subsubsection{Optically thick case}

If we assume that the HDO line shape is determined by the line opacity
rather than pure kinematics, then the line asymmetry is due to
blueshifted absorption, revealing again outflowing gas (because an optically thick infalling envelope would lead to redshifted absorption). 
We can obtain stringent constraints on the
size, density, temperature, and HDO abundance of the emitting 
gas.
To this end, in the non-LTE LVG predictions in Fig. 4 
we add dashed curves where the HDO line opacity is equal to
unity, i.e. we assume a moderate thickness to explain the line absorption.  
Also, the lower level of the HDO transition is
very high ($E_{\rm l}$= 318 K): 
the upper panel of Fig. 4 shows that if the source size is
0$\farcs$3, the  $\tau$ = 1 curves (dashed) are always shifted
to the right of the  curves of the observed signal (solid) and never overlap: 
in other words, any optically thick layer of this size (or larger) 
would emit more in HDO than observed in our beam.
In order to avoid excessive emission, i.e. to have the
two families of curves intersect,  
and to constrain that 
$N$(HDO)/$N$(H$_2$) does not exceed the 10\% of the D/H elemental abundance, 
one needs to assume a larger beam dilution factor
requiring a source size
smaller than 0$\farcs$08 (37 AU) and larger than 0$\farcs$04 (18 AU). 
The lower panel of
Fig. 4 shows the intermediate case, with 0$\farcs$06 (28 AU). 
The solution is, in
this case, for temperatures between 70 K and 110 K and
densities of $10^9$--$10^{10}$ cm$^{-3}$. 
In particular, for a temperature around 100 K,
$N$(HDO)$\sim 7 \times 10^{17}$ cm$^{-2}$, corresponding to 
$N$(HDO)/$N$(H$_2$)=$1.7\times 10^{-6}$--$10^{-7}$. Assuming
a typical water abundance of $\sim 10^{-4}$ would imply
HDO/H$_2$O $\sim 1.7\times 10^{-2}$--$10^{-3}$, consistent with the
values measured in hot corinos (Persson et al. 2014; Ceccarelli et al. 2015).
The optically thick case thus also requires slow outflowing gas,
further constraining the physical conditions, i.e. a size $\simeq$ 18--37 AU,
$T_{\rm kin}$ $\simeq$ 100 K, and $n_{\rm H_2}$ $\geq$ 10$^9$ cm$^{-3}$.

\begin{figure}
\centerline{\includegraphics[angle=0,width=7.5cm]{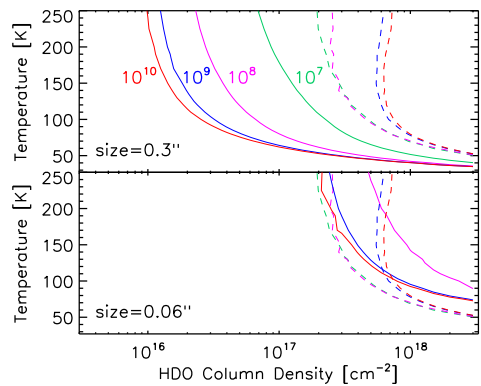}}
\caption{LVG predictions of the temperature versus HDO column density required to
reproduce the observed velocity-integrated emission (solid curves) and
to have unit line opacity (dashed curves) for 
densities of: 10$^{7}$ cm$^{-3}$ (green), 10$^{8}$ cm$^{-3}$ (magenta), 
10$^{9}$ cm$^{-3}$ (blue), and
10$^{10}$ cm$^{-3}$ (red). Optical depth increases with column density.
The upper panel refers to a source size of
0$\farcs$3 and the lower panel to 0$\farcs$06.}
\label{hdolvg}
\end{figure}

\section{Conclusions}

The combination of high sensitivity, high angular resolution, and large
bandwidth offered by ALMA has allowed us to image  high-excitation ($E_{\rm u}$ up to 335 K) 
CH$_3$CHO and HDO emission in the inner $\simeq$ 100 AU of the
Sun-like HH212 protostar for the first
time.  
Both HDO and CH$_3$CHO emission 
indicates $T_{\rm kin}$ larger than 70 K, while HDO
requires gas
densities $\geq$ 10$^{7}$ cm$^{-3}$. 
We thus report the detection of the first hot corino in Orion.
The acetaldehyde abundance is similar to that 
measured in hot corinos located
in low-mass star forming regions in Ophiuchus. 

The asymmetric HDO profile at low velocities  
indicates that at least some of the deuterated water
is associated with slow outflowing gas, where a 
high SO abundance (up to 10$^{-7}$) has also been detected
by Podio et al. (2015).  
These finding support a chemical enrichment of
the low-velocity outflowing gas heated by
the protostar in its surroundings.
If the emission is optically thick 
the emitting size must be very small (18--37 AU) and the density 
has to be extremely high, with $n_{\rm H_2}$ $\geq$ 10$^9$ cm$^{-3}$.
With such extreme sizes and densities, it is tempting to  
speculate that the observed gas may be associated with
a disk wind gas accelerated at the base.
Interestingly, the occurrence of a wide-angle flow in HH212 
with a nested onion-like velocity structure
has  recently been suggested by C$^{34}$S observations 
with ALMA (Codella et al. 2014). 

\begin{acknowledgements}
We acknowledge the anonymous referee for instructive comments.
This paper makes use of the ADS/JAO.ALMA\#2011.0.000647.S data (PI: C. Codella). ALMA is a partnership of 
ESO (representing its member states), NSF (USA), and NINS (Japan), together with NRC (Canada) and NSC 
and ASIAA (Taiwan), in cooperation with the Republic of Chile. The Joint ALMA Observatory is operated by ESO, AUI/NRAO, and NAOJ. 
This work was partly supported by the PRIN INAF 2012 -- JEDI and by the 
Italian Ministero dell'Istruzione, Universit\`a e Ricerca through the grant Progetti Premiali 2012 -- iALMA.
LP has received funding from the European Union Seventh Framework Programme (FP7/2007-2013) under grant agreement n. 267251.
\end{acknowledgements}

\clearpage

\appendix

\section{The CH$_3$CHO rotational diagram}

Table 1 lists the emission lines
observed towards HH212--MM1 and used for the
standard analysis of the rotational
diagram (Fig. A.1), which allows us to derive $T_{\rm rot}$ = 87$\pm$47 K
and $N_{\rm tot}$ = 2$\pm$1 $\times$ 10$^{15}$ cm$^{-2}$.

\begin{figure}
\centerline{\includegraphics[angle=0,width=9cm]{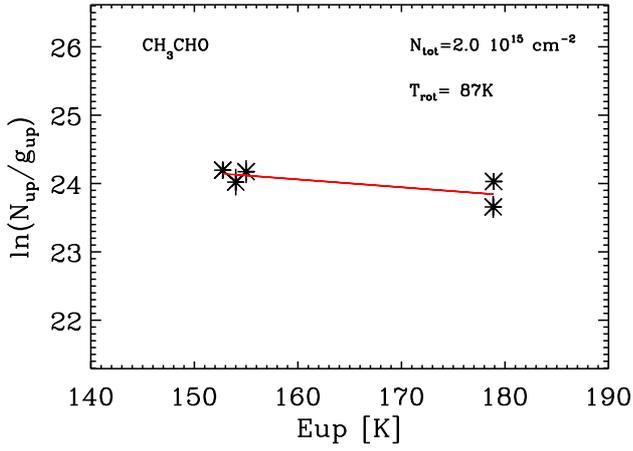}}
\caption{Rotation diagram for the 
CH$_3$CHO transitions reported in Table 1. The parameters
$N_{\rm up}$, $g_{\rm up}$, and $E_{\rm up}$ are respectively
the column density, the degeneracy, and the energy of the upper level.
We note that the error bars on ln($N_{\rm up}$/$g_{\rm up}$) are
given by the vertical bar of the symbols.
The $g_{\rm up}$ value is 74 for all the transitions used here 
(and reported in Table 1) with the
exception of   17$_{\rm 2,15}$--16$_{\rm 2,14}$ A, where  $g_{\rm up}$ = 70.  
The plot allows us to derive a rotational temperature of
87$\pm$47 K and a total column density of 2$\pm$1 $\times$ 10$^{15}$ cm$^{-2}$.} 
\label{rdplot}
\end{figure}

\end{document}